\def\year{2017}\relax
\documentclass[letterpaper]{article}
\usepackage{aaai17}
\usepackage{times}
\usepackage{helvet}
\usepackage{courier}
\usepackage{amsmath}
\usepackage{xcolor}
\usepackage{graphicx}
\usepackage{subfigure}
\graphicspath{{Images/}}
\usepackage{float}
\usepackage{times}
\usepackage{url}
\usepackage{latexsym}
\usepackage{amsmath}
\usepackage{graphicx}
\usepackage{color}
\usepackage{pgf}	
\usepackage{multicol}
\usepackage{lipsum}
\usepackage{enumerate}
\usepackage{xspace}
\usepackage[numbers]{natbib}
\usepackage{booktabs}

\newcommand{\xhdr}[1]{\paragraph*{\bf #1}}
\newcommand{\omt}[1]{}

\begin{document}
\title{Tracing the Use of Practices through Networks of Collaboration}
\author{Rahmtin Rotabi \\
Department of Computer Science \\
Cornell University \\
Ithaca, NY, 14853 \\
\texttt{rahmtin@cs.cornell.edu}
\And Cristian Danescu-Niculescu-Mizil  \\
Department of Information Science \\
Cornell University \\
Ithaca, NY, 14853 \\
\texttt{cristian@cs.cornell.edu}
\And Jon Kleinberg\\
Department of Computer Science \\
Cornell University \\
Ithaca, NY, 14853\\
\texttt{kleinber@cs.cornell.edu}
}

\maketitle
\begin{abstract}
An active line of research has used on-line data to study the
ways in which discrete units of information---including messages, photos,
product recommendations, group invitations---spread through
social networks.
There is relatively little understanding, however, of how on-line data
might help in studying the diffusion of more complex {\em practices}---roughly, routines or styles of work that are generally handed down
from one person to another through collaboration or mentorship.
In this work, we propose a framework together with a novel type of data 
analysis that seeks to study the spread of such practices by tracking 
their syntactic signatures
  in large document collections.
Central to this framework is the notion of an {\em inheritance graph}
that represents how people pass the practice on to others 
through collaboration.
Our analysis of these inheritance graphs demonstrates that we can
trace a significant number of practices over long time-spans,
and we show that the structure of these graphs can help in predicting
the longevity 
of collaborations within a field,
as well as the fitness of the practices themselves.
\end{abstract}

\section{Introduction}

On-line domains have provided a rich collection of settings in which
to observe how new ideas and innovations spread through social networks.
A growing line of research has discovered principles for both the
local mechanisms and global properties involved in the spread
of pieces of information such as messages, quotes, links, news stories, and photos
\cite{adar-blogspace,gruhl-blogspace,leskovec-blogspace-sdm07,liben-nowell-pnas08,leskovec-kdd09,adamic-how-met-me,cheng-www14,goel-structural-virality,barbieri2013topic},
the diffusion of new products through viral marketing
\cite{leskovec-ec06j}, 
and the cascading recruitment to on-line groups
\cite{backstrom-kdd06,anderson-www15}.

A common feature in these approaches has been to trace some discrete
``unit of transmission'' that can be feasibly tracked through the underlying 
system: a piece of text, a link, a product, or membership in a group.
This is natural: the power of on-line data for analyzing diffusion
comes in part through
the large scale and fine-grained resolution with which 
we can observe things flowing
 through a network;
therefore, to harness this power it is crucial for those things to be algorithmically recognizable and trackable.
As a result, certain types of social diffusion have been particularly
difficult to approach using on-line data---notably,
a broad set of cascading behaviors that we could refer to
as {\em practices}, which are
a collection of styles or routines
within a  community that are passed down between people over many years,
often through direct collaboration, mentorship or instruction.
Particular stylistic elements involved in writing software, 
or performing music, or playing football, might all be examples
of such practices in their respective fields.
While complex practices are one of the primary modes studied by
qualitative research in diffusion \cite{strang-diffusion},
the challenge for large-scale
quantitative analysis has been both to recognize when someone has
begun to adopt a practice, and also to identify how it was transmitted to them.

\xhdr{Tracking the Spread of Practices.}
A natural approach to tracking the spread of a practice is to 
find a concretely recognizable ``tag'' that tends to travel with
the practice as it is handed down from one person to another, 
rendering its use and transmission easily visible.
A beautiful instance of this strategy was carried out by 
David Kaiser in his analysis of the use of {\em Feynman diagrams} in physics
\cite{kaiser-feynman-diagrams}.
Feynman diagrams were proposed by Richard Feynman as a way to 
organize complex physics calculations, and due to the technical 
sophistication involved in their use, the initial spread of
Feynman diagrams within the physics community proceeded in much 
the style described above, with young researchers adopting the
practice through collaboration with colleagues who had already used it.
In contrast to many comparable practices, Feynman diagrams had a
distinctive syntactic format that made it easy to tell when they
were being used.  As a result, their spread could be very accurately
tracked through the physics literature of the mid-20th-century.
The result, in Kaiser's analysis, was a detailed map of how an idea spread
through the field via networks of mentorship. As he writes:
\begin{quote}
{\footnotesize \em
The story of the spread of Feynman diagrams reveals the work
required to craft both research tools and the tool users who will put
them to work. The great majority of physicists who used the diagrams
during the decade after their introduction did so only after working
closely with a member of the diagrammatic network. Postdocs
circulated through the Institute for Advanced Study, participating in
intense study sessions and collaborative calculations while there.
Then they took jobs throughout the United States (and elsewhere) and
began to drill their own students in how to use the diagrams. To an
overwhelming degree, physicists who remained outside this rapidly
expanding network did not pick up the diagrams for their research.
Personal contact and individual mentoring remained the diagrams'
predominant means of circulation even years after explicit
instructions for the diagrams' use had been in print.  \cite{kaiser-feynman-diagrams}
} 
	\end{quote}

The Feynman diagram thus functions in two roles in this analysis:
as an important technical innovation, and as a ``tracking device'' for 
mapping pathways of mentorship and collaboration.
If we want to bring this idea to a setting with large-scale data, 
we must deal with the following question: where can we find a 
rich collection of such tracking devices with which to perform this
type of analysis?  We do not expect most objects in this
collection to be technical advances comparable to the Feynman diagram,
but we need a large supply of them, and we need 
to be able to mechanically recognize both their use and their spread.

\xhdr{The present work: Diffusion of practices in academic writing.}
In this paper, we describe a framework for tracking the spread of
practices as they are passed down through networks of collaboration,
and we demonstrate a number of ways in which our analysis has
predictive value for the underlying system.
We make use of a setting where practices have the recognizability that
we need---a novel dataset of latex macros in the e-print arXiv 
recently developed by Rotabi et al \cite{rotabi-www17}.
While the earlier work that developed this dataset
used macros for other purposes (specifically,
treating macros names as instances of naming conventions),
macros in our context have a number of the key properties we need.
First, a latex macro is something whose presence can be tracked
as it spreads through the papers in the arXiv collection;
we can thus see when an author first uses it, 
and when their co-authors use it.
Second, while an arbitrary macro clearly does not 
correspond in general to an important technical innovation, a sufficiently
complex macro often does encode some non-trivial technical shorthand within
a concrete sub-field, and hence its use signifies the corresponding
use of some technical practice within the field.
And finally, there are several hundred thousand latex macros in 
papers on the arXiv, and so we have the ability to track a huge number
of such diffusion events, and to make comparative statements about
their properties.

\begin{figure*}[ht]
\centering
\subfigure[\texttt{\textbackslash hbox\{\$\textbackslash  rm\textbackslash thinspace L\_\{\textbackslash odot\}\$\}}]{\label{fig: Barcons}\includegraphics[width=\linewidth]{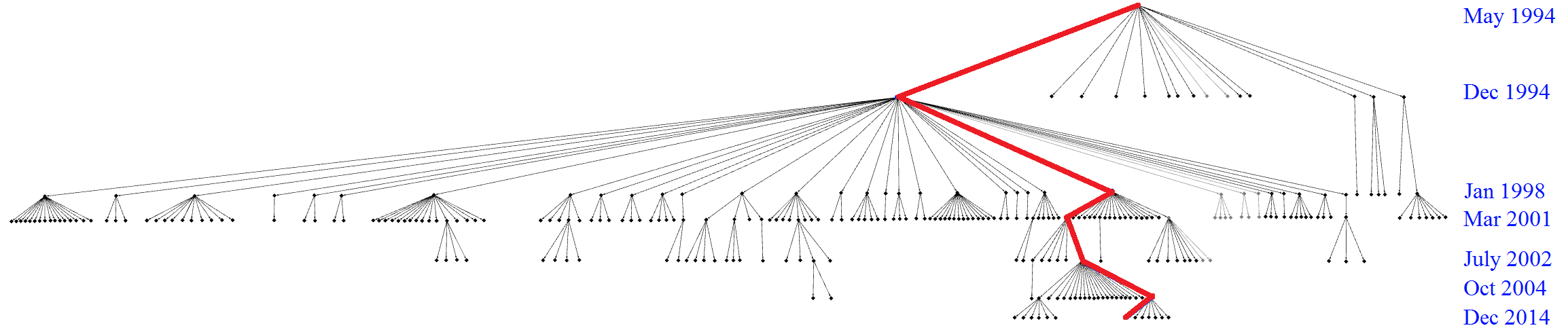}}
\subfigure[\texttt{\textbackslash mbox\{\textbackslash boldmath \$Y\$\}}]{\label{fig: Ng}\includegraphics[width=0.48\linewidth]{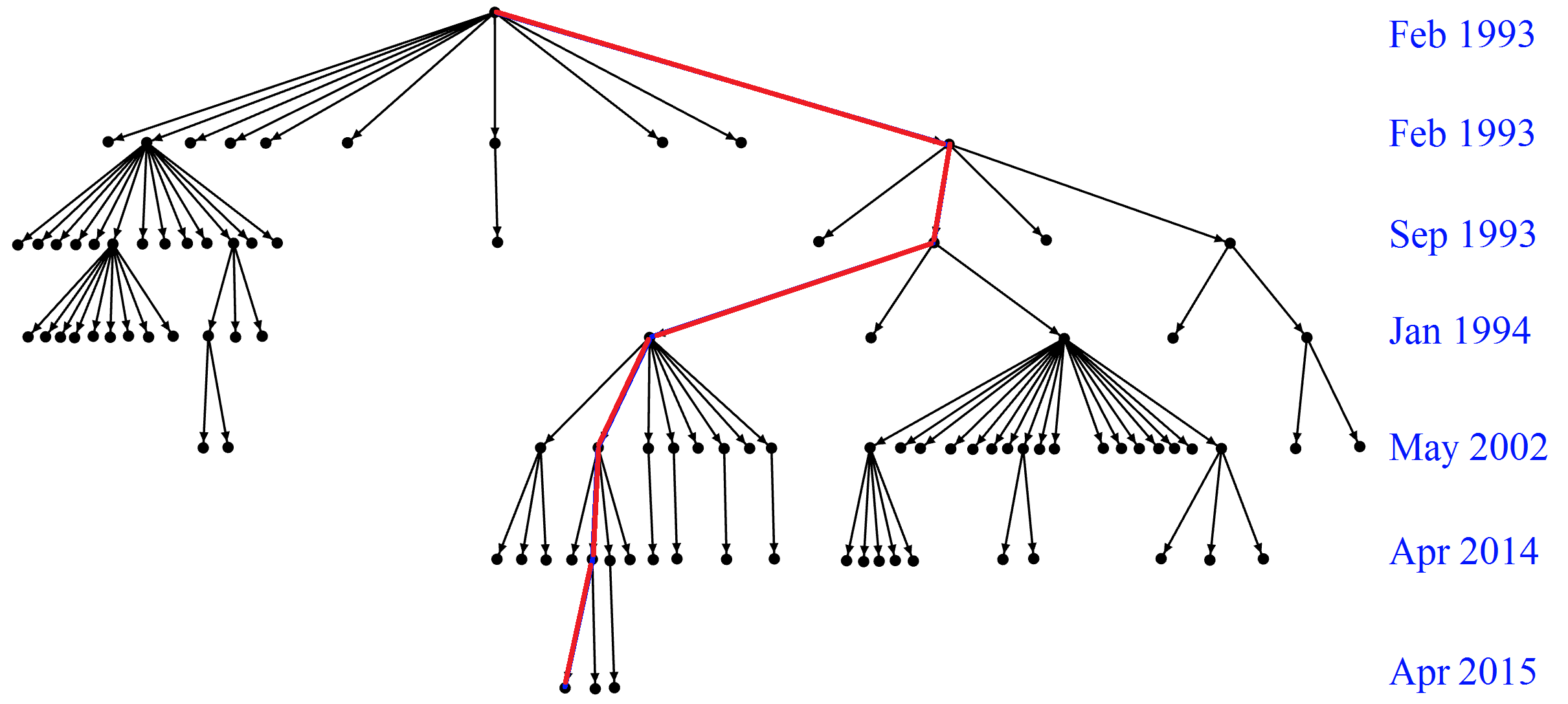}}
\subfigure[\texttt{\textbackslash mathrel\{\textbackslash mathpalette\textbackslash@versim$>$\}}]{\label{fig: Lopez}\includegraphics[width=0.48\linewidth]{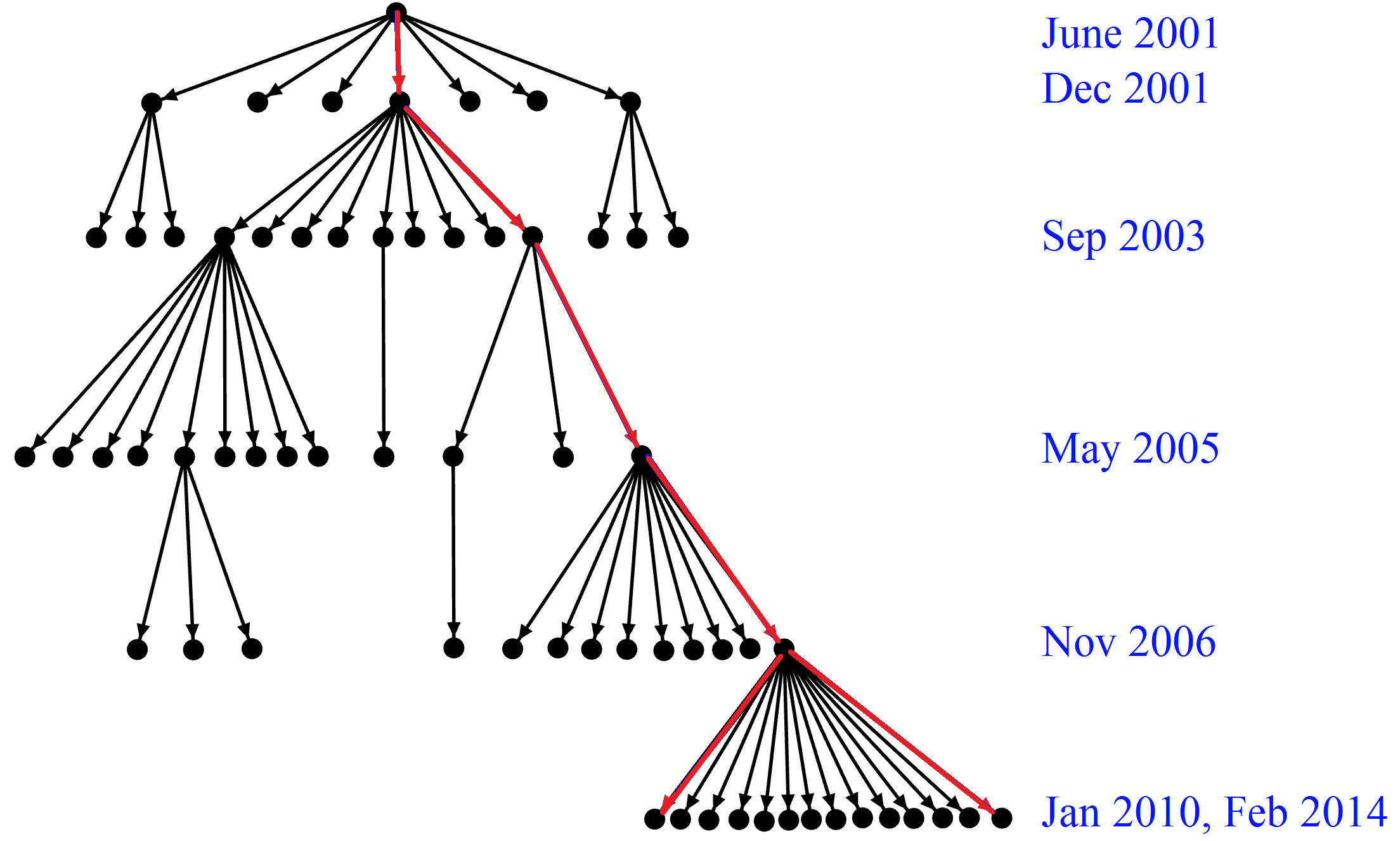}}
\caption{Sample subsets of BFS trees for three different macros. At each depth we show the date when the highlighted author (node) uses the macro for the first time; the highlighted edge is a paper in which an author passes on the macro to an author (node) in the next level of the tree.}
\label{fig: Trees}
\end{figure*}

If we want to use macros to trace the diffusion of practices between
collaborators, we first need to establish whether macros 
indeed spread via ``inheritance''
from co-authors: as with the Feynman diagram, 
can most of the initial set of uses of a macro trace a path back to 
a single early use through a chain of co-authorship?  
We find that this is true for a significant
fraction of macros, by using an {\em inheritance graph} for each macro
that records how each author's first use can be imputed to a co-authorship
with an earlier user of the macro.  
Specifically, for each macro we can build a graph on the set of authors
who have used it, and we include a directed {\em inheritance edge}
from author $u$ to author $v$ if (i) $u$ used the macro before $v$ did,
and (ii) $v$'s first use of the macro is in a paper with $u$.
We find that many of these
inheritance graphs contain giant directed subtrees rooted 
at a single early use of the macro, indicating that a significant fraction of
the users of the macro can indeed 
trace a direct path back to a single shared early ancestor
under this inheritance relation.

These structures represent interesting instances of diffusion for
several reasons.  First, they are ``organic'' in a way that
the spread of many on-line memes are not: when we study 
on-line diffusion in settings where a user's exposure to content is governed
by a recommendation system or ranking algorithm, there is the
added complexity that part of the diffusion process is being guided
by the internals of the algorithms underlying the system.
With macros in arXiv papers, on the other hand,
while authors may use automated tools
to format the source of their papers, there is relatively little
influence from automated recommendations or rankings in the
actual decisions to include specific macros.
Second, we are studying processes here that play out over 
years and even decades; among other findings about the structure of
our inheritance graphs, we observe that their diameters can take
multiple years to increase even by one hop.
We are thus observing effects that are taking place over multiple
academic generations.

\xhdr{The present work: Estimating fitness.}
If these inheritance graphs---obtaining by tracing simple syntactic signatures
in the source files of papers---are telling us something about the
spread of practices through the underlying community, then their 
structural properties may contain latent signals about the outcomes
of authors, topics, and relationships.
In the latter part of the paper, we show that this is the case,
by identifying such signals built
from the inheritance structures, and showing that they have predictive value.

As one instance, suppose we wish to estimate the future longevity
of a collaboration between two authors $u$ and $v$---that is,
controlling for the number of papers they have written thus far,
we ask how many papers they will write in the future.
If $(u,v)$ is an edge of the inheritance graph for some macro, does
this help in performing such an estimate?  
One might posit that since this edge represents something concrete
that $u$ passed on to $v$ in their collaboration, we should 
increase our estimate of the
strength of the relationship and hence its future longevity.
This intuition turns out not to be correct on its own:
the existence of a $(u,v)$ edge by itself doesn't significantly modify 
the estimate.  
However, we find that something close to this intuition does apply.
First, we note that since a $(u,v)$ edge only means that a macro used by
$u$ showed up subsequently in a paper that $u$ co-authored with $v$,
it is providing only very weak information about $v$'s role in the interaction.
We would have a stronger signal if $(u,v)$ were an {\em internal edge}
of some inheritance graph, meaning that $v$ has at least one outgoing edge;
in this case, $v$ was part of a paper that subsequently passed the macro
on to a third party $w$.
We find that if $(u,v)$ is an internal edge of an inheritance graph,
this does in fact provide a non-trivial predictive signal for 
increased longevity of the $u$-$v$ collaboration;
informally, it is not enough that $u$ passed something on to $v$,
but that $v$ subsequently was part of the process of
passing it on to a third party $w$.
In fact, we find something more: when $(u,v)$ is an edge that is not
internal (so that $u$'s passing on of the macro ``ends'' at $v$),
it in fact provides a weak predictive signal that the collaboration
will actually have slightly {\em lower} longevity than an
arbitrary collaboration between two co-authors (again controlling
for the number of joint papers up to the point of observation).

In what follows, we formalize this analysis and its conclusion.
We also develop analyses through which macro inheritance can be used to
help estimate the future longevity of an author---how many papers 
will they write in the future? ---and the fitness of an individual macro 
itself---how many authors will use it in the future?

The remainder of the paper is organized into three main sections.
We first briefly describe the structure of the data and how it is
used in our analyses.
We then formally define the inheritance graphs and survey some
of their basic properties.
Finally, we analyze the relation between these inheritance structures and the longevity of co-authorships, authors, and macros.

\section{Data Description}
\label{section:Data}
\newcommand{\name}{\texttt{name}\xspace}
\newcommand{\names}{\texttt{names}\xspace}
\newcommand{\Name}{\texttt{Name}\xspace}
\newcommand{\Names}{\texttt{Names}\xspace}
\newcommand{\body}{\texttt{body}\xspace}
\newcommand{\bodies}{\texttt{bodies}\xspace}

The dataset we study contains the macros used in over one million
papers submitted to the e-print arXiv from its inception in 1991 through 
November 2015. The arXiv is a repository of scientific
pre-prints in different formats, primarily in \LaTeX. 
Macros have two major components, the \name and the \body. Whenever the
author uses \textbackslash\name the \LaTeX \xspace compiler replaces it with
the \body and compiles the text. In our study the \body serves as the
``tracking device'' discussed in the introduction, for studying
how a macro is passed between collaborators over time.
In general, when we refer to a ``macro'', we mean a macro \body
unless specified otherwise.
For our study we use macro \bodies that have length greater than 20 
characters, and which have been used by at least 30 different authors.
We apply the length filter so that we can focus on macros that
are distinctive enough that we expect them to move primarily through
copying and transmission, rather than independent invention.
Further information can be
found in \cite{rotabi-www17}, which introduces this dataset.

\section{Method}

\section{Inheritance Graphs}
\xhdr{Defining inheritance graphs.}
\begin{figure*}[th]
\centering
\subfigure[]{\label{fig: Biggest_Cascade_Fraction_CDF}\includegraphics[width=0.31\linewidth]{{{Feynman/BiggestComponentFraction_CDF}}}}
\subfigure[]{\label{fig: max_path}\includegraphics[width=0.31\linewidth]{{{Feynman/AverageTimeDepth}}}}
\subfigure[]{\label{fig: onion_shape}\includegraphics[width=0.31\linewidth]{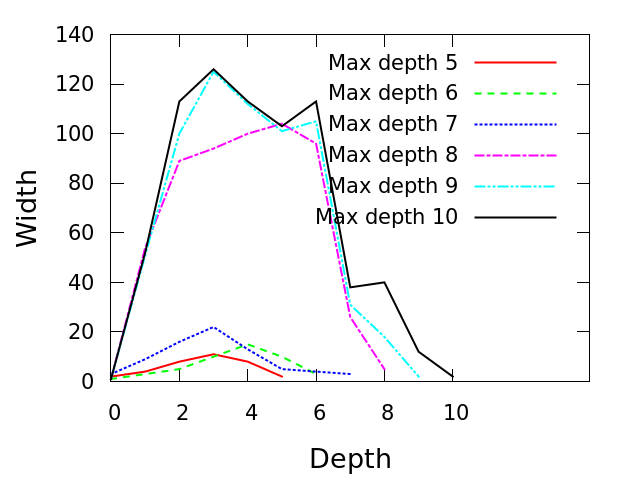}}
\caption{(a) The CDF for the ratio of the largest reachable set to the number of nodes in the graph. (b) The average number of months that pass from the date of appearance of the root paper to the date of appearance of nodes at a given depth, grouped by the maximum depth of the tree. (c) The average number of nodes in each depth, for the largest reachable set for each macro.}
\label{fig: Components}
\end{figure*}
We begin by formally defining the {\em inheritance graphs} described
in the introduction.
For each macro
$m$ we create a graph $(V_m, E_m)$ where $V_m$ is the set of
authors who have used macro $m$ in at least one of their papers. 
We add a directed edge $(u,v)$ to the edge set $E_m$ if 
there is a paper that uses $m$ with $u$ and $v$ as co-authors, such that
(i) this is $v$'s first use of $m$, but 
(ii) $u$ has used $m$ in at least one previous paper.
This is the formal sense in which 
$m$ is being passed from $u$ to $v$:
$v$'s first use of $m$ occurs in collaboration with $u$,
a prior user of $m$. 
Note that there can be multiple edges leading into a single node.
For instance take a paper with authors $u,v$ and $z$ that uses macro $m$, and
 assume that $u$ and $v$ have used $m$ before but $z$ is using it 
for the first time; then both the edges $(u,z)$ and $(v,z)$ are in the graph.

Now, if all authors of a paper $p$ are using $m$ for the first time,
then the nodes corresponding to these authors will not have any
incoming edges. (Nodes of this form are the only ones with no
incoming edges.)  For each such paper $p$, we replace the nodes
corresponding to the authors of $p$ with a single {\em supernode}
corresponding to $p$.
We will refer to this as a {\em source node}, and to the authors
of $p$ as {\em source authors}.
The resulting graph, with supernodes for papers where no author has
used the macro before, and with author nodes for all others, is
the inheritance graph $G_m$ for the macro $m$.
Because the process of inheritance, as defined, goes forward
in time, $G_m$ is necessarily a directed acyclic graph (DAG).

Using these graphs we should be able to trace back a macro's life to
its inception and to the authors who first used it. 
Note that there might be multiple source papers, and hence several
groups of co-authors who independently serve as ``origins'' for the macro.
For portions of the analysis where we are interested in looking at the number
of authors who all follow from a single source paper, we will identify
the source paper that has directed paths to the largest number of nodes in the 
graph $G_m$.  We will refer to this as the {\em seed paper}, and
to the set of authors of this paper as the {\em seed authors}.
(Note that the seed paper might not be the chronologically earliest
paper to use the macro $m$; it is simply the one that can reach the
most other nodes.)

\xhdr{Analyzing the inheritance graphs.}
Our dataset contains several hundred thousand different macros, and
as a first step we analyze
the properties of the graphs $G_m$ that they produce. 
In Figure \ref{fig: Trees} we take three sample macros and show subsets of
the breadth-first search (BFS) trees that are obtained starting from the 
seed paper.
For example in Figure \ref{fig: Barcons} the graph is created on the
macro,  \texttt{\textbackslash hbox\{\$\textbackslash rm\textbackslash
thinspace L\_\{\textbackslash odot\}\$\}} and the seed node is the
paper {\textbf{astroph/9405052}} with authors Xavier Barcons and
Maria Teresa Ceballos.  The seed paper used this macro in 1994, and
some of the nodes at depth 6 in the BFS tree are from 2014---a 20-year
time span to reach a depth of 6 in the cascading adoption of the macro.
This reinforces the sense in which we are studying cascades that
play out on a multi-generational time scale of decades, rather than
the time scale of minutes or hours that characterizes many on-line cascades.
The seed node 
of Figure \ref{fig: Ng} is the paper {\textbf{hep-th/0106008}} with
authors Selena Ng and Malcolm Perry, and the seed node of Figure \ref{fig:
Lopez} is the paper {\textbf{hep-ph/9302234}} with authors Jose R
Lopez et al.  Since all other nodes in these BFS trees have incoming
edges, they all correspond to individual authors who enter the graph
at their first adoption of the macro, whereas the root node corresponds
to a single paper and to the contracted set of authors of this paper.

We now consider some of the basic properties of these inheritance graphs.
First, each source paper has a {\em reachable set} in $G_m$---the set of nodes
it can reach by directed paths---and recall that we defined the
seed paper to be the source paper with the largest reachable set.
In Figure \ref{fig: Biggest_Cascade_Fraction_CDF} we observe that a
non-trivial fraction of the macros have a seed paper whose reachable
set is a large fraction of all the authors who eventually adopt the macro.
This provides a first concrete sense in which the inheritance patterns
contained in $G_m$ represent a global structure that spans much of the
use of the macro $m$.

In Figure \ref{fig: max_path} and \ref{fig: onion_shape} we show the properties of the graphs and
nodes grouped based on the maximum depth of the BFS tree and the 
depth of the individual nodes.
Figure \ref{fig: max_path} shows the average time it takes for the
macro to get from the root to the nodes in each depth grouped by the
maximum depth of the tree. This figure shows how these cascades
can take multiple years to add a single level of depth to the tree,
and a decade or more to reach their eventual maximum depth.
In Figure \ref{fig: onion_shape} we show the median width (number
of nodes) of trees at each depth, again grouped by the maximum depth
of the tree.  Based on this plot we see that most of these trees have
are narrow in their top and bottom layers, with fewer nodes, and
are wider in the middle.

The plots thus far have been concerned with the global structure of
the inheritance graph and its shortest paths as represented by 
breadth-first search trees.
Now we take a deeper look at the properties of individual edges in the graph.
For this we will first define the notions of local and global
experience, and we will use these two terms throughout the paper. 
At time $t$ the {\em{global experience}} of an author is the number of
papers the respective author has written. At time $t$ the {\em{local
experience}} of an author is defined with respect to a macro $m$ and
is the number of papers up to time $t$ in which the author has used $m$.
This is a version of the notion of local experience relative to an
arbitrary term, as used in \cite{rotabi-icwsm16}.

\begin{figure}[h]
\centering
\includegraphics[width=0.48\textwidth]{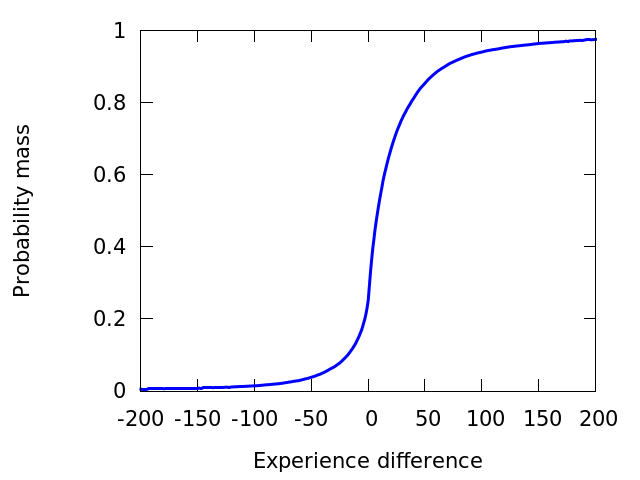}
\caption{The Cumulative Distribution Function of the global experience difference between the source and destination of an edge.}
\label{fig: Teacher-Learner}
\end{figure}
Consider an edge $(u,v)$ in the inheritance graph for a macro $m$.
At the moment when the macro is passed from $u$ to $v$, the local experience
of $v$ with respect to $m$ is 0 by definition, and the local experience of
$u$ with respect to $m$ is greater than 0.
What do we expect about the global experience of these two nodes?
To the extent that passing on a macro is a form of ``teaching''
from one person to another, we may expect the global experience of
$u$ (the ``teacher'') to be higher than the global experience of $v$
(the ``learner'').  On the other hand, there is a history of 
sociological work in the diffusion of innovations suggesting that
innovations often originate with outsiders 
who come from the periphery of the system
\cite{danescu-www13,mclaughlin-change-agent-revisited,simmel-sociology,valente-network-interventions},
which would be consistent with $v$ having higher global experience than $u$.
Figure \ref{fig: Teacher-Learner} addresses this question by
showing the cumulative distribution
of the global experience difference between $u$ and $v$.
The median experience difference is clearly shifted in the positive
direction, consistent with the ``teacher'' node $u$ having the
higher global experience in general.

\section{Fitness}
\label{sec:fitness}

Now that we have some insight into how the information diffusion
process unfolds in our data, we investigate whether these
inheritance structures can provide predictive signal for the outcomes
of co-authorships, authors, and the macros themselves.
In all cases we will think in terms of the {\em fitness}
of the object in question---the extent to which it survives
for a long period of time and/or produces many descendants.

\subsection{Fitness of collaborations}

We start by considering the fitness of collaborations---given two
authors $u$ and $v$ who have written a certain number of papers 
up to a given point in time, or perhaps who have not yet collaborated,
can we use anything in the structure
of macro inheritance to help predict how many more papers they will
write in the future?

A natural hypothesis is that if $v$ inherits macros from $u$, then
this indicates a certain strength to the relationship (following
the teacher-learner intuition above), and this may be predictive
of a longer future history of collaboration.
To examine this hypothesis, we perform the following computational test 
as a controlled paired comparison.
We find pairs of co-authorships $u$-$v$ and $u'$-$v'$ with properties
that (i) neither pair has collaborated before, (ii) their first co-authorship
happens in the same month, (iii) $(u,v)$ is an edge in an inheritance
graph, and (iv) $(u',v')$ is not.
(Note that since we are looking at pairs of co-authorships, we
are looking at four authors in total for each instance: $u$, $v$, $u'$,
and $v'$.)
Now we can ask, aggregating over many such pairs of co-authorships,
whether there is a significant difference in the future number of papers
that these pairs of authors write together.
(Since their initial co-authorships took place in the same month, they
have a comparable future time span in which to write further papers.)

In fact, we find that there isn't a significant difference, at odds
with our initial hypothesis about macro inheritance.
However, there is more going on in the inheritance structure that
we can take advantage of.
We divide the edges of the inheritance graphs into two sets:
{\em internal edges} $(u,v)$, where the node $v$ has at least
one outgoing edge, and {\em terminal edges} $(u,v)$, where
the node $v$ has no outgoing edge.
Internal edges add extra structural information, since they 
indicate that 
not only $u$ passed the macro $m$ to $v$, but that $v$ was then part of the process of passing $m$ in a collaboration
subsequent to the one in which they originally inherited it.

\begin{figure}[!ht]
     \includegraphics[width=\linewidth]{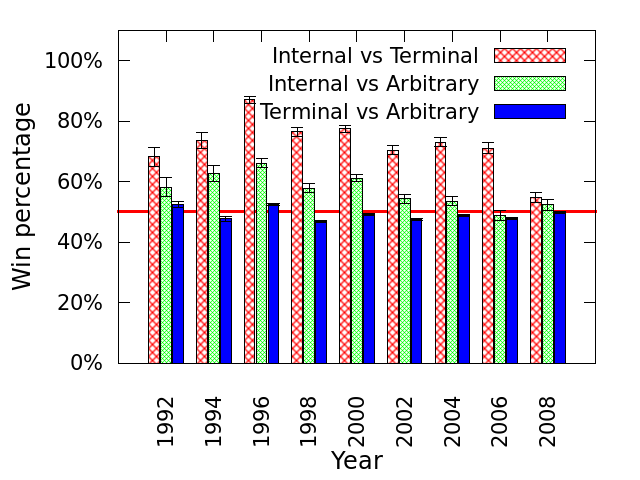}
	\caption{Comparison of three different co-authorship settings through different years in the data. The bars show the win percentage of the first of the two listed categories;
	 e.g., the red bar indicates the percentage of times co-authors with internal edges end up writing  more papers than the matched co-authors with terminal edges.
  The horizontal red line indicates the 50\% baseline.}

\label{fig: CoAuthorshipFitness}
\end{figure}	

We find that the fitness of $u$-$v$ co-authorships is significantly
higher when $(u,v)$ forms an internal edge,
in contrast to the lack of effect when $(u,v)$ is an arbitrary edge.
We evaluate this using an extension of our previous paired comparison:
in conditions (i)-(iv) above for forming pairs of co-authorships,
we replace conditions (iii) and (iv) with the following:
\begin{itemize}
\item Internal edge vs. arbitrary co-authorship:
(iii) $(u,v)$ is an internal edge and (iv) $(u',v')$ is not an edge.
\item Internal edge vs. terminal edge:
(iii) $(u,v)$ is an internal edge and (iv) $(u',v')$ is a terminal edge.
\item Terminal edge vs. arbitrary co-authorship:
(iii) $(u,v)$ is a terminal edge and (iv) $(u',v')$ is not an edge.
\end{itemize}
In each of these three settings, we look at the fraction of times
that one of the categories produced the co-authorship with more 
future papers.  In our paired setting, if we were to draw two co-authorships
uniformly at random over all possible co-authorships (without
regard to the type of the edge),
there is a 50\% chance that the first would produce the higher
number of future papers.
Thus,  we can calibrate each of the three comparisons
listed above using this 50\% baseline.
Figure \ref{fig: CoAuthorshipFitness} shows these results,
grouped into two-year bins: we find that internal edges win a
large fraction of the comparisons against each of the other two categories, 
whereas there is little difference between terminal edges and
arbitrary co-authorships.

\subsection{Fitness of authors}

We now consider the fitness of the authors themselves; we will show that
the way authors use macros can provide a weak but non-trivial signal
about how many papers they will eventually write, a quantity
that we refer to as the {\em fitness} of the author.

The particular property we consider is a type of ``stability''
in the usage of the macro.
For a given macro \body, there are many possible \names that
can be used for it, and authors differ in the extent to which 
their papers preserve a relatively stable choice of names for the same
macro body: some almost always use the same name, while for other
authors the name changes frequently.
(For example, an author who almost always uses the name  
\texttt{\textbackslash{vbar}} for the macro body
\texttt{\$\textbackslash overline\{v\}\$},
versus an author whose papers alternate between using
\texttt{\textbackslash{vbar}}, \texttt{\textbackslash{barv}},
\texttt{\textbackslash{vb}}, \texttt{\textbackslash{vbarsymb}},
and others, all for this same macro body.)
We could think of the first type of author as exerting
more control over the 
source of her papers than the second type of author, 
and this distinction between the two types of authors---based
on their behavior with respect to macros---naturally raises the
 question whether the stability of macro names
could provide predictive value for author fitness.

Here is how we formally define this measure.
For a particular author $a$, we say they change
the name of macro $m$ on paper $p$ if the previous time they used $m$'s
macro body, the name was different.
Then, for a set of authors $A$ and a set of macros $M$, we define
$f(A,M,x)$ to be the probability of an author in $A$ changing the name
of a macro in $M$ the $x^{\rm th}$ time 
they use it. We consider  this {\em name-change probability}, $f(A,M,x)$ for $x \in [0,40]$
and different groups of authors and macros. In particular we look at
groups of authors that have more than $\theta$ papers in the entire
corpus. We set $\theta$ to be $40, 50, \ldots, 130$ and we let $M$ 
range over three possible sets: the set of all macros;
the set of {\em wide-spread macros} (more than 250
authors use the macro body); and the set of {\em narrow-spread macros} 
(at least 20
authors used it and at most 250). 

One source of variability in this analysis is that even once we
fix the minimum number of papers $\theta$ written by an author $a$,
as well as the usage number $x$ of the macro $m$ that we are considering,
it is still possible that author $a$'s $x^{\rm th}$ use of the macro
might come toward the end of their professional lifetime or
early in their professional lifetime.
(It must come at the $x^{\rm th}$ paper they write or later,
since they need time to have used the macro $m$ a total of $x$ times,
but this is all we know.)
It is easy to believe that authors who use a macro in their early
life stages might exhibit different phenomena from those who use it
in a later life stage.
Therefore, in addition to the measures defined so far,
we also consider analyses involving only the set of macro uses that 
come early in the authors' professional lifetime---specifically
only macro uses that happen in the author's first 40 papers.

\begin{figure}[t!]
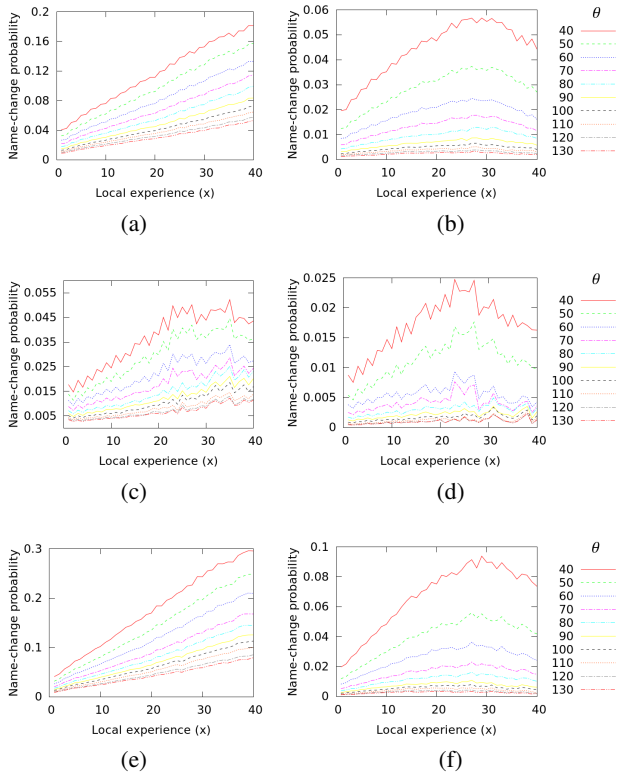

\centering
\subfigure[]{\label{fig:All-All}\includegraphics[width=0.43\linewidth]{{{PeopleChangeOverTime/LocalExperience_Change_WithGlobalThresholds-All-All}}}}
\subfigure[]{\label{fig:Early-All}\includegraphics[width=0.55\linewidth]{{{PeopleChangeOverTime/LocalExperience_Change_WithGlobalThresholds-EarlyLife-All}}}}

\subfigure[]{\label{fig:All-narrow}\includegraphics[width=0.43\linewidth]{{{PeopleChangeOverTime/LocalExperience_Change_WithGlobalThresholds-All-narrow}}}}
\subfigure[]{\label{fig:Early-narrow}\includegraphics[width=0.55\linewidth]{{{PeopleChangeOverTime/LocalExperience_Change_WithGlobalThresholds-EarlyLife-narrow}}}}

\subfigure[]{\label{fig:All-wide}\includegraphics[width=0.43\linewidth]{{{PeopleChangeOverTime/LocalExperience_Change_WithGlobalThresholds-All-wide}}}}
\subfigure[]{\label{fig:Early-wide}\includegraphics[width=0.55	\linewidth]{{{PeopleChangeOverTime/LocalExperience_Change_WithGlobalThresholds-EarlyLife-wide}}}}

\caption{Each panel shows the probability an author changes the name of
a macro on their $x^{\rm th}$ use of it.  A single curve in each plot
shows the set of all authors with at least $\theta$ papers, for 
$\theta$ equal to $40, 50, \ldots, 130$.
Each row of panels corresponds to a different set of macros:
the first row shows results for the set of all macros; the second for
the set of narrow-spread macros; and the third for the set of
wide-spread macros (as defined in the text).
The left column of panels shows the analysis for each of these three sets
over the authors' full professional lifetimes.
The right column of panels shows the analysis for each of these three sets
restricted to the authors' early life stages (first 40 papers only).
Thus, the panels are
 (a) full lifetimes, all macros; (b) early life stages, all macros;
 (c) full lifetimes, narrow-spread macros; (d) early life stages, 
 narrow-spread macros; (e) full lifetimes, wide-spread macros;
 (f) early life stages, wide-spread macros.
}
\label{fig:ColorPlotsAllMacros}

\end{figure}

The results for all these settings are shown in Figure
\ref{fig:ColorPlotsAllMacros}: the three  sets of macros
(all macros, wide-spread macros, and narrow-spread macros); 
for each of these sets, we consider both the authors' full lifetimes
and just their early life stages.
In each case, 
the x-axis shows the number of macro uses (i.e. the authors'
local experience with respect to the macro), and the different
curves represent authors grouped by different values of the
minimum number of papers $\theta$.

This suggests that overtime authors build a certain ``loyalty'' to the names
they have used consistently; this is consistent with our previous findings regarding
the competition between macro-naming conventions\cite{rotabi-www17}.

But we also find something else: that (eventually) more prolific authors (larger
$\theta$) have a lower name-change probability (compare ordering of curves in 
each subplot of Figure \ref{fig:ColorPlotsAllMacros}.
This suggests that the macro name change probability might be a signal
with predictive value for author fitness (which, again, we define
as the number of papers the author will eventually write).

To test this idea, we set up an author fitness prediction task as follows.
For a given minimum number of papers $\theta$ we consider the low-fitness authors to be the ones with fitness below the $20^{th}$ percentile and high-fitness
authors to be those above the $80^{th}$ percentile.
We then see whether simply using the frequency with which an author
changes macro names in the first $\theta$ papers can serve as a predictor for this two-class problem:
whether an author's fitness is below the 20th percentile or above
the 80th percentile.

\begin{table}
\centering
\begin{tabular}{c c c}
\toprule
Papers revealed ($\theta$) & 20'th Percentile & 80'th Percentile \\ 
\midrule
10 & 13 & 38 \\ 
20 & 25 & 58 \\ 
30 & 36 & 73 \\ 
40 & 47 & 87 \\ 
50 & 58 & 99 \\ 
\bottomrule
\end{tabular} 
\caption{Global experience thresholds used in defining the author fitness 
classes for each number $\theta$ of papers	revealed.}
\label{table:20_80_percentile}	
\end{table}

\begin{figure}[!ht]
     \includegraphics[width=\linewidth]{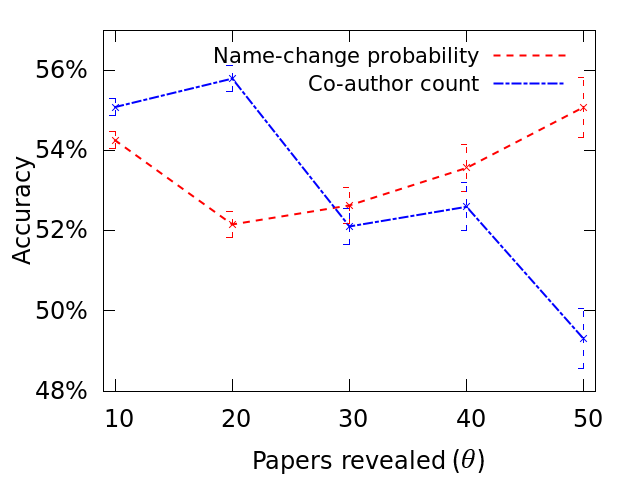}
	\caption{The accuracy of predicting the number of publications of an author given her first few papers, $\theta$. We compare the performance of the name-change
	probability features with the features based on number of co-authors.}
	\label{fig: AuthorFitness}
\end{figure}

By using the probability of macro name changes, we are able to predict
which of these two classes an author belongs to with a performance
that exceeds the random baseline of $50\%$ by a small but statistically significant amount.
Figure \ref{fig: AuthorFitness} shows the performance for different values
of $\theta$.
We emphasize that predicting an author's fitness is a challenging
task for which one doesn't expect strong performance even from
rich feature sets; this makes it all the more striking
that one can obtain non-trivial performance from 
the frequency of macro name changes, a very low-level property about
the production of the papers themselves.

Moreover, for settings involving large values of $\theta$ the name-change probability is more predictive than an arguably more natural structural feature: the author's total number of co-authors (Figure \ref{fig: AuthorFitness}).  We also note that in such settings the name-change feature also outperforms other more direct macro-based features, such as the total number of macros used, or total number of distinct macro bodies used.

\subsection{Fitness of macros}

Finally, we consider the fitness of the macros themselves.
We define the fitness of a macro to be the total number of authors
who eventually use the macro body, and investigate which features
are predictive of this variable.

We set up a prediction task as follows. We first find all macros that
get adopted by at least $k$ authors.  Each of these macros has a fitness
(of at least $k$), and we define $\sigma(k)$ to be the median
of this multiset of fitness values: of all macros that reach at least $k$ authors,
half of them have a fitness of at most $\sigma(k)$, and half of them 
have a fitness of at least $\sigma(k)$.
In table \ref{table: MacroFitness} we report
$\sigma(k)$ and the number of macro instances for different values of $k$.

\begin{table}[ht]
\centering
\begin{tabular}{c c c c}
\hline 
\toprule
$k$ & $\sigma(k)$ &  Instances \\ 
\midrule
40  & 98  & 49,415 \\ 
80 & 156 & 30,107 \\ 
120 & 242 & 20,119 \\ 
160 & 340 & 14,662 \\ 
200 & 437 &  11,794 \\ 
\bottomrule
\end{tabular}
\caption{Summary of the macro fitness prediction dataset:
For a macro that reaches at least $k$ authors, the task is to predict whether
it will eventually reach $\sigma(k)$ authors (the median fitness of such
macros). }
\label{table: MacroFitness}
\end{table}

We can thus use $\sigma(k)$ to construct a balanced prediction task,
in the style of the cascade prediction analyses from \cite{cheng-www14}.
For a given macro that reaches at least $k$ authors, 
we observe all the information on the papers and authors up to the
point at which the $k^{\rm th}$ author adopts the macro, 
and the task is then to predict if this macro will eventually reach $\sigma(k)$ authors.
We learn a logistic regression model
for different values of $k$ and report the accuracy in Figure \ref{fig:
Macro_Fitness} on an $80$-$20$ train-test split.\footnote{We can achieve a $1\%$
to $4\%$ better accuracy by using a non-linear classifier such as
decision trees, but we opt to use the more interpretable model.}

\begin{figure}[!ht]
     \includegraphics[width=\linewidth]{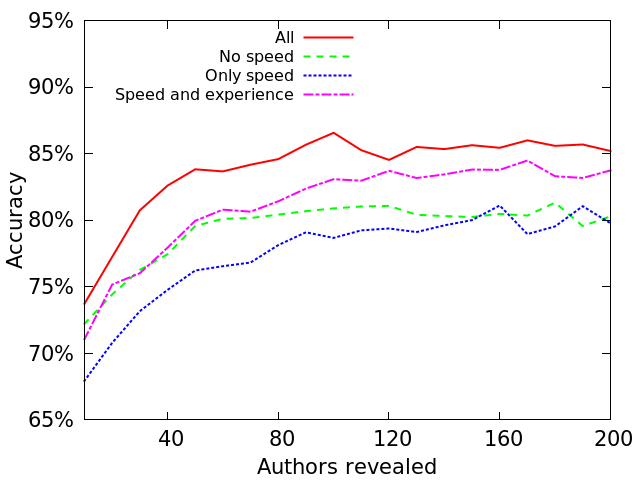}
	\caption{The accuracy of predicting how widely a macro spreads,
using different subsets of features.}
	\label{fig: Macro_Fitness}
\end{figure}

We use the following features.
\begin{itemize}
\item Features related to the {\em diffusion speed} of the macro: 
the number of papers that the macro
needs in order to reach $\frac{k}{2}$ and $k$ distinct authors;
and the number of months that the macro
needs in order to reach $\frac{k}{2}$ and $k$ distinct authors.
\item {\em Experience of the macro users}:
the average usage experience of the first $k$ authors who adopted it.
\item {\em Structural features of the macro users}:
the local and global
clustering coefficients of the co-authorship graph on the first
$k$ authors to use the macro.
\item {\em Structural features of the macro body}:
the length of the macro body, the number of dollar
signs in the macro 
(generally used for mathematical notation), the number of non-alphanumerical characters, and the
maximum depth of nested curly brackets.
\end{itemize}

In Figure \ref{fig: Macro_Fitness} we show the prediction 
performance for different subsets of these features, as a function of $k$;
note that performance increases with increasing $k$.
As observed above, predicting macro fitness is a problem whose
syntactic form is closely analogous to the prediction of cascade
size for memes in social media \cite{cheng-www14};
given this, and the fact that the spread of macros plays out over
so much longer time scales, and without the role of ranking or
recommendation algorithms, it is interesting to note 
the similarities and contrasts in the prediction results.
One of the most intriguing contrasts is in the role of diffusion
speed features: for cascade prediction in social media,
the speed features alone yielded performance almost matching
that of the full feature set, and significantly outperforming
the set of all non-speed features\cite{cheng-www14}.
For our domain, on the other hand, the speed features perform
5-10\% worse than the full feature set; they also perform
worse for most values of $k$ than the set of all non-speed features.
This suggests that for macro fitness, the speed features 
are considerably less powerful than they are in the social media context,
indicating that there may be more to be gained from the synthesis
of a much broader set of features.

\section{Conclusions}

The spread of practices between collaborators is a challenging
form of diffusion to track, since one needs to be able to 
recognize when someone has begun using a practice, and how it
was conveyed to them.  Motivated by work that used the Feynman
diagram as an easily recognizable ``tracer'' of a complex
practice \cite{kaiser-feynman-diagrams}, we track the spread
of several hundred thousand macros through the papers of the
e-print arXiv over a 25-year period.
Long macros often serve as technical shorthand within a
defined sub-field, and their syntactic precision makes it easy
to follow their flow through the collaboration network.
We construct {\em inheritance graphs} showing how the macro 
spread between collaborators, and 
we find that many macros have a clear ``seed set'' of authors
with the property that a large fraction of the subsequent users of the macro
can trace a direct inheritance path back to this seed set.
The resulting diffusion patterns are intriguing, in that they
span multiple academic generations and several decades, and unlike
cascades in social media, the spread of these macros takes 
place with very little influence from ranking or recommendation algorithms.

We also find that properties of macro inheritance provide signals
that are predictive for larger-scale properties that have nothing
to do with macros.  These include predictions about the longevity of 
collaborations and the number of papers that an author will write
over their professional lifetime on the arXiv.

Our work suggests a number of directions for future research.
First, it would be interesting to develop a comparative analysis
between the structure of our inheritance graphs and the 
corresponding structures for the diffusion of on-line memes.
Are there systematic ways in which the two types of diffusion patterns
differ, and can these be connected to differences in the
underlying mechanisms?
Second, we believe that there may well be additional links between
inheritance structures and prediction problems for the trajectory of
the overall system; for example, can we evaluate the future course
of larger sub-areas based on the inheritance patterns that exhibit?
And finally, identifying ``tracers'' for complex practices is
a style of analysis that can applied in other domains as well;
as we broaden the set of contexts in which we can perform this type
of analysis, we may better understand the ways in which the flow
of practices helps reinforce and illuminate our understanding of
large collaborative communities.

\xhdr{Acknowledgments} 
We are grateful to Paul Ginsparg for his valuable advice and for his help with
the arXiv dataset. We would like to thank Peter Lepage for introducing
us to David Kaiser's analysis of the diffusion of Feynman diagrams, which 
formed the basis for our thinking about the approach in this paper.
This work was supported in part by ARO, Facebook, Google, the Simons 
Foundation, and a Discovery and Innovation Research Seed Award from
Cornell's OVPR.

\bibliographystyle{abbrv}
\bibliography{refs}
\end{document}